\documentclass[12pt]{iopart}

\usepackage{graphicx}  
\begin{document}

\title{Fourier transform spectrometry without Fourier analysis of the
interferogram}

\author{Miguel Lagos$^1$, Rodrigo Paredes$^2$ and C{\'e}sar Retamal$^3$}
\address{$^1$ Departamento de Tecnolog{\'\i}as Industriales, Facultad de
Ingenier{\'\i}a, Universidad de Talca, Campus Los Niches, Camino a Los
Niches km 1, Curic{\'o}, Chile.}
\address{$^2$ Departamento de Ciencias de la Computaci{\'o}n, Facultad de
Ingenier{\'\i}a, Universidad de Talca, Campus Los Niches, Camino a Los
Niches km 1, Curic{\'o}, Chile.}
\address{$^3$ Departamento de Ciencias de la Construcci{\'o}n, Facultad de
Ingenier{\'\i}a, Universidad de Talca, Campus Los Niches, Camino a Los
Niches km 1, Curic{\'o}, Chile.}
\ead{mlagos@utalca.cl}

\vspace{10pt}

\begin{abstract}
It is shown here that precision is gained by analyzing the
interferometric spectra directly from the interferograms, with no
previous Fourier transformation to put them in the standard frequency
domain. The method is based on the theoretical calculation of the
lineshape, which gives a general closed--form expression for the
spectrum in the time domain and is directly assimilable to the
experimental interferogram. Error sources in Fourier integrals,
apodization, peak fitting with standard interpolation functions,
choice of background level, Stokes energy shifts, etc., are neatly
avoided.
\end{abstract}

\vspace{2pc}
{\it Keywords}: optical spectra, FTIR spectroscopy, quantum optics,
chemical analysis

%
%
%

\section{Introduction}

We show here the scientific basis for an alternate way to process the
raw data constituting the output of Fourier transform spectrometers,
which is expected to improve substantially the precision of both
concentration and energy measurements. In a standard traditional
dispersive spectrometer, good frequency resolution is attained at the
cost of blocking by the narrow slit of the monochromator most of the
photons that otherwise would reach the detector. This brings down the
signal to noise ratio. A Fourier transform spectrometer replaces the
monochromator by a Michelson interferometer which yields the cosine
Fourier transform of the spectrum as the output. The procedure
collects information at all frequencies simultaneously (multiplex
advantage), improving dramatically both speed and signal to noise ratio
because the detector captures the full intensity of the light coming
from the sample (throughput advantage). The output is an interferogram
consisting on a graph giving the radiation intensity as a function of
the difference in the optical path length of the two arms of the
interferometer, which is accurately measured from the interference
fringes of a reference laser (Connes advantage). It is referred to as
the raw data, and usually exhibits a complex oscillating structure
which must be Fourier transformed to bring out the spectrum in
conventional way. The spectrometer then has to be associated to a
numerical processor to display the spectrum. Infrared (IR) absorption
spectroscopy went through a major advance when Fourier transform
spectroscopy (FTIR) came to the fore, and practically no dispersive IR
equipment is in the market today. However the procedure is also
practical in optical spectroscopy, nuclear magnetic resonance
spectroscopy and magnetic resonance spectroscopic imaging
\cite{Griffiths,Stuart,Hlivko}.

However, the numerical integration of strongly oscillating functions
may give large errors and hence performing the Fourier transform of
the raw data to display the spectrum in conventional way may be not a
minor task. By this reason, the feasibility of the new technique was
associated to the development of a mathematical method for the fast
calculation of the Fourier transform of highly structured functions
(the FFT algorithm). The discovery of this method by Cooley and Tukey
in 1965 initiated a new generation of IR instruments and techniques
\cite{CooleyTukey}. Anyway, processing of the raw data to display a
standard spectrum involves some error. Beside the numeric errors, it
is also necessary to control spurious spectral features created by the
truncation of the interferogram constituting the raw data. Experimental
scans are necessarily finite, and the sudden cutoffs at the boundaries
have broad Fourier representations which must be recognized and
discarded by a procedure known as apodization. Hence the Fourier 
transform of the raw data constitutes the first main source of error.

The other source of error is the little attention usual methods gives
to the precise physical origin of line shapes, broadenings and shifts. 
Precise quantitative chemical analysis and the accurate determination
of the excitation energies of molecular bonds demand more elaborate
mathematical processing of the spectra, additional to the Fourier
transformation of the raw data. The concentration of a chemical
species is determined by the area under the peak identifying a
characteristic bond of it, whose evaluation demands curve fitting of
the data, particularly in the presence of heavy overlapping or
structured background. Gauss, Lorentz and Voigt distributions are in
practice the analytical expressions used for fitting the shape of the
spectral peaks and determining their areas by integration. However,
these distributions are merely interpolation functions because, in
rigor, do not follow from solving a real physical model for the
processes causing the peak broadenings \cite{Dodd}. Gauss and Lorentz
curves are analytically very different, especially concerning the
contribution of the tails to the peak area, which is much more
significant for the latter. Tails immerse in the noisy background and
their effect owns to the domain of experimental uncertainty when using
a tentative model for the lineshape. It has been proven that the
assigned peak intensities may show substantial variations with the
choice of the lineshape model \cite{Marshall}. The symmetry of these
standard distributions evidences their limited ability to describe
spectral profiles. It has been demonstrated on a general basis that
the lineshapes for photon absorption and emission by atomic or
molecular species in a condensed environment are always asymmetric
with respect to the net energy of the electronic transition
\cite{Lagos1,Lagos2}. Hence a second main source of error is the
adoption of a standard distribution not well grounded on the physics
of the target to describe the spectral features. The fit of the
experimental data by the mathematical curves given by a realistic
model for the profiles of the spectral features greatly improves
precision. If the theoretical curve reproduces well the physics of the
radiation field interacting with the target then the error is given by
the statistical dispersion of the experimental points along the curve
that bests the fit, instead of its whole breadth
\cite{Lagos1,Lagos2,LagosParedes1,LagosParedes2,LagosParedesRetamal}.

\section{Lineshape functions of electromagnetic spectra}

Line broadenings and energy shifts are produced mainly by multiphonon
processes involving the extended acoustic modes of vibration of the
condensed medium embedding the photosensitive orbital. They are
activated by the local distortion that follows the sudden excitation
or de--excitation of the electronic bonding orbitals, and can be
calculated analytically, yielding a closed--form mathematical
expression for the lineshape function \cite{Lagos1,Lagos2}. It is
fortunate that this theoretical expression for the spectral
distribution in the standard frequency domain has the general form of
the Fourier transform of a function in the time domain. This way, the
theory gives directly what experimentalists call the {\it raw data}
and there is no need in principle of calculating the Fourier
transforms of the theoretical and experimental results to compare them
and make them to fit. The first and second main sources of error are
thus avoided by making the theoretical analysis of the spectra in the
time domain, working directly with the raw data.

The lineshape function has been proven in the recent literature to be
given by the integral expression
\cite{Lagos1,Lagos2,LagosParedes1,{LagosParedesRetamal}}

\begin{eqnarray}
\fl F(\hbar ck;T)
=\frac{a}{\pi\hbar v_s}\int_{-\infty}^\infty\, d\tau\,
\bigg\{\exp\big[ -\alpha J(\tau;T)\big]
-\exp\big[ -\alpha J(\infty ;T)\big]\bigg\} \nonumber\\
\times\exp\bigg\{ i\bigg[\alpha I(\tau )-
\frac{2a}{\hbar v_s}(\hbar ck-E)\tau\bigg]\bigg\}
+\exp\big[-\alpha J(\infty ;T)\big]\,\delta (\hbar ck-E),
\label{E1}
\end{eqnarray}

\noindent
where $\hbar ck$ is the photon energy, $E$ the energy difference of
the two electronic states involved in the transition, $a$ is
essentially the bond length, and $v_s$ the mean speed of sound of the
acoustic modes of vibration of the medium. The adimensional constants
$\alpha$ and $\beta$, and the adimensional dummy time $\tau$ are given
by

\begin{equation}
\alpha =\frac{3(\Delta F)^2}{\pi^2\hbar\rho v_s^3}
\quad\beta =\frac{\hbar v_s}{2 a k_B T}
\quad\tau =\frac{v_s}{2a}t ,
\label{E2}
\end{equation}

\noindent
where $\Delta F$ is the bond mean force variation upon excitation,
$\rho$ is the density, $k_B$ is the Boltzmann constant, and $T$ the
temperature. The auxiliary functions $J(\tau;T)$ and $I(\tau )$ are
dependent on the symmetry of the surroundings of the orbital
undergoing the transition. For the simplest case of octahedral
coordination (OC) of the optically active orbital they read

\begin{equation}
J(\tau; T)=
\int_0^{aq_D}\,\frac {dx}{x}\, \bigg( 1-\frac{\sin x}{x}\bigg)
\coth (\beta x)\sin^2 (\tau x) \quad \rm{(OC)}
\label{E3}
\end{equation}

\begin{equation}
I(\tau )=\frac{1}{2}
\int_0^{aq_D}\,\frac {dx}{x}\,\bigg( 1-\frac{\sin x}{x}\bigg)
\sin (2\tau x)\quad \rm{(OC)},
\label{E4}
\end{equation}

\noindent
with $q_D$ being the Debye wavevector of the acoustic waves,
$aq_D=(12\pi^2)^{1/3}$, and

\begin{equation}
J(\infty ; T)=\frac{1}{2}
\int_0^{aq_D}\,\frac {dx}{x}\,\bigg( 1-\frac{\sin x}{x}\bigg)
\coth (\beta x) \quad \rm{(OC)}.
\label{E5}
\end{equation}

The second term in the right hand side of Eq.~(\ref{E1}) for $F(\hbar
ck;T)$, containing the delta--function, is the zero--phonon line, and
the first one is the phonon broadened distribution. The lineshape
function $F(\hbar ck;T)$ is normalized as
\cite{LagosParedes1,Lagos1,Lagos2}

\begin{equation}
\int_{-\infty}^{\infty} d(\hbar ck)\, F(\hbar ck;T)=1
\label{E6}
\end{equation}

\noindent
and hence the relative contribution of zero--phonon processes to the
total is $I_{ZPL}=\exp\big[-\alpha J(\infty ;T)\big]$.

Other symmetries may give more complicated functional forms for $J$
and $I$. For example, for tetrahedral coordination (TC) of the
optically active orbitals one has that

\begin{equation}
J(\tau; T)=\int_0^{aq_D}\,\frac{dx}{x}\,\bigg[\frac{5}{2}
-\frac{3}{2}\frac{\sin x}{x}
-\frac{\sin\bigg(\frac{1}{2}\sqrt{3}x\bigg)}{\frac{1}{2}\sqrt{3}x}
\bigg]\coth (\beta x)\sin^2 (\tau x)\quad\rm{(TC)},
\label{E7}
\end{equation}

\begin{equation}
I(\tau )=\frac{1}{2}
\int_0^{aq_D}\,\frac {dx}{x}\,\bigg[\frac{5}{2}
-\frac{3}{2}\frac{\sin x}{x}
-\frac{\sin\bigg(\frac{1}{2}\sqrt{3}x\bigg)}{\frac{1}{2}\sqrt{3}x}
\bigg]\sin (2\tau x) \qquad\rm{(TC)}.
\label{E8}
\end{equation}

\begin{figure}[h!]
\begin{center}
\includegraphics[width=9cm]{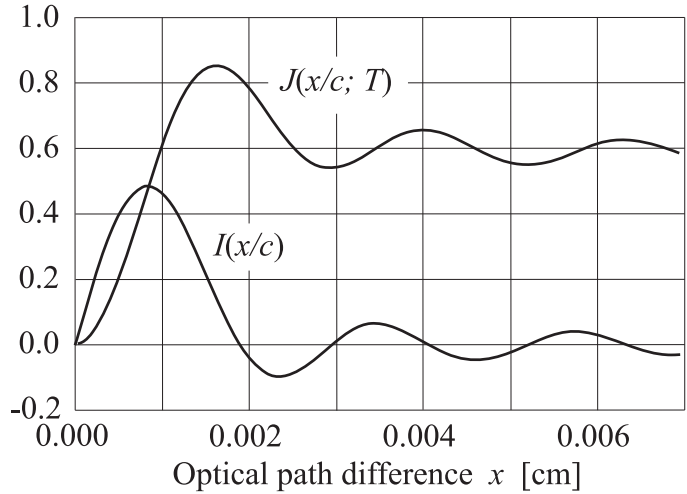}
\caption{\label{Fig1} The auxiliary functions $J(x/c;T)$ and $I(x/c)$
defined in Eqs.~(\ref{E3}) and (\ref{E4}) for octahedral symmetry with
$\tau =x/c$.}
\end{center}
\end{figure}

The output $f(x,T)$ of the Michelson interferometer, where $x$ is the
difference in the optical path lengths of the two arms of the
interferometer, is the cosine Fourier transform of the spectrum given
by the lineshape function (\ref{E1}), i.~e.

\begin{equation}
f(x;T)=hc\int_{-\infty}^{\infty} d\bar{k}\, F(hc\bar{k};T)
[1+\cos{(2\pi i\bar{k}x)}],
\label{E9}
\end{equation}

\noindent
where $\bar{k}=1/\lambda$ is the wave number and $h$ the Planck
constant. Substituting Eq.~(\ref{E1}) and solving the integral this
gives

\begin{equation}
f(x;T)=1+
\exp\bigg[ -\alpha J\bigg(\frac{x}{c};T\bigg)\bigg]
\cos\bigg[\alpha I\bigg(\frac{x}{c}
\bigg)+\frac{Ex}{\hbar c}\bigg],
\label{E10}
\end{equation}

\noindent
where $x$ in fact plays the role of the virtual time $t=x/c$,
conjugate to the angular frequency $\omega=E/\hbar$. Function $f(x;T)$
is actually the interferogram which FTIR spectroscopists call the {\it
raw data}. Therefore, replacing in Eq.~(\ref{E10}) the pairs of
auxiliary functions (\ref{E3}) and (\ref{E4}) for octahedral symmetry
of the optically active orbitals, or (\ref{E7}) and (\ref{E8}) for
tetrahedral coordination of them, one obtains explicit closed--form
mathematical expressions for the interferograms. Auxiliary functions
for other symmetries can be derived from the general expressions for
the electron--phonon coefficients, given in Ref.~\cite{Lagos2}.

\section{The proposed method}

As both the experimental technique and the general theory, which is
well grounded on the physics of the energy transfers between the
radiation field and the charges in a condensed system, arrive both to
the interferogram expressed by Eq.~(\ref{E10}), in principle there is
no need to perform any Fourier transform of the data to grasp the
physical information from the experimental results. The analytical
closed--form expression (\ref{E10}) depends on only a few parameters,
$\alpha$, $\beta$ and the net transition energy $E$, per spectral line.
Hence the most practical way to proceed is to find the constants
$\alpha$, $\beta$ and $E$ by means of a best fit analysis of
Eq.~(\ref{E10}) to the experimental interferograms. Fourier analysis
then becomes just an optional alternative for people which likes to
identify spectra in the conventional frequency domain. 

The method seems highly convenient because retains all the advantages
of Fourier spectrometry avoiding the numerical errors associated to
the Fourier integration of rapidly oscillating functions. An important
example of this is given by the area under the spectral line, which in
agreement with Eq.~(\ref{E9}) is given by half the intensity
$f(0,T)/2$ of the central maximum of the interferogram. This magnitude
is unity under the hypothesis of a single emission or absorption
center of our theory, but in empirical grounds is proportional to the
number of optically active orbitals and gives the concentration of
them. However, the implementation of practical procedures for
interpreting the measured interferograms directly with Eq.~(\ref{E10})
may be not immediate, particularly when dealing with narrow spectral
lines, precisely because their interferograms oscillate strongly with
$x$. In particular, the central maximum may be very narrow and its
intensity may be strongly affected by the experimental uncertainty of
$x$.

\begin{figure}[h!]
\begin{center}
\includegraphics[width=10cm]{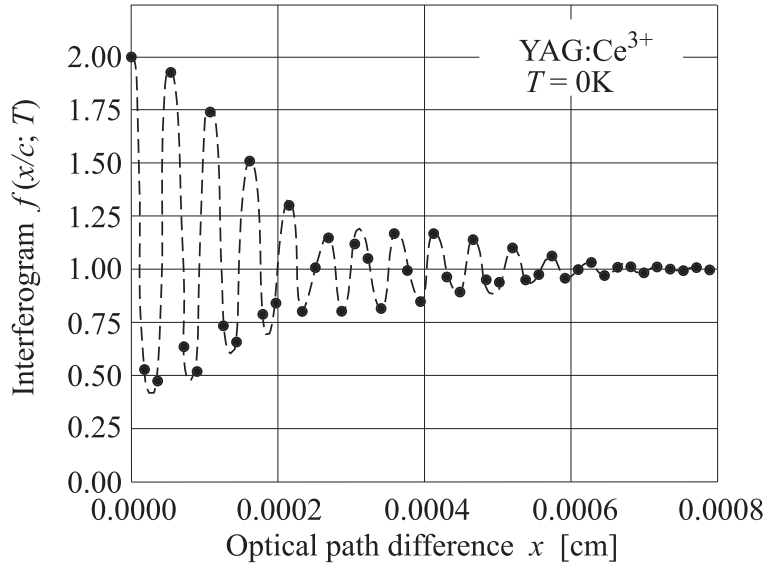}
\caption{\label{Fig2} Filled black circles represent the function
$f(x;T)$ given by Eq.~(\ref{E11}), where $f_1$ and $f_2$ have the
general form (\ref{E10}) with slightly different parameters $\alpha$
and $E$. It is expected that $f(x;T)$ will reproduce the experimentally
registered Michelson interferogram of two partially resolved spectral
features emitted by centers which have concentrations in the ratio 2:1.
The broken line is only a guide to the eye.}
\end{center}
\end{figure}

\begin{figure}[h!]
\begin{center}
\includegraphics[width=11cm]{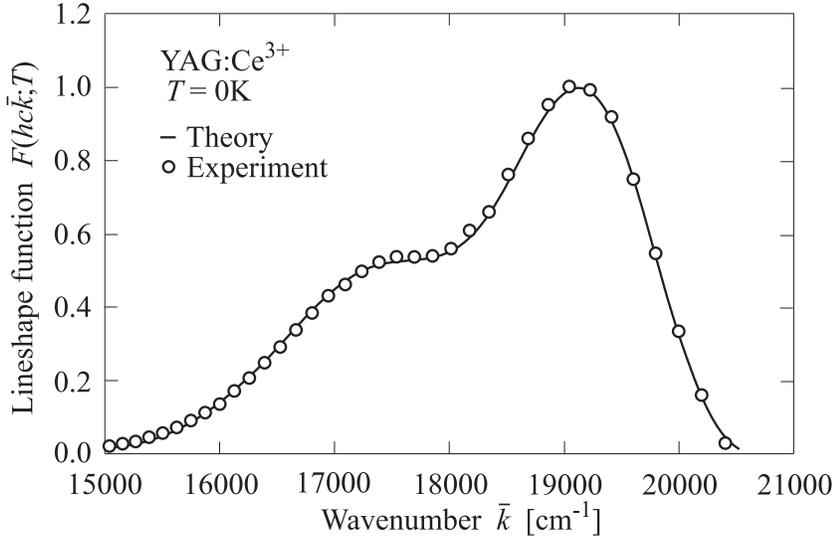}
\caption{\label{Fig3} Open circles represent the fluorescent emission
spectrum of YAG:Ce$^{3+}$ at $T=4\,\rm{K}$, as measured by Bauchmann
et al. [1]. The solid line represents the Fourier transform of the
function represented in Fig.~\ref{Fig2} by the discrete set of filled
black circles.}
\end{center}
\end{figure}

The general procedure can be applied to both wide spectral features,
like those displayed by fluorescent compounds, or narrow ones, as the
sharp minima observed in the absorption spectra of infrared light
passing through many materials. The physical process is essentially
the same. Figures \ref{Fig1}, \ref{Fig2} and \ref{Fig3} show the
mathematical steps of the calculation of the spectrum of YAG:Ce$^{3+}$
(yttrium aluminum garnet, $\rm{Y}_3\rm{Al}_5\rm{O}_{12}$, doped
with Ce$^{3+}$), a phosphor having many technical applications. This
system is particularly interesting because its emission spectrum has
been measured at a temperature close to $T=0$ with a resolution large
enough to clearly observe at $\lambda =489\,\rm{nm}$
($\bar{k}=20450\, {\rm{cm}^{-1}}$) the zero--phonon line belonging
to the main of the two emission bands \cite{Bachmann}. The relative
intensity of the zero--phonon line of YAG:Ce$^{3+}$ at temperature
$T=4\,\rm{K}$ is observed to be 0.27\% of the total intensity of the
main emission band. The emission has been attributed to competing
transitions \cite{Bachmann,LagosParedesRetamal} of the
$\rm{Al\,O}_6$ groups of quasi--octahedral coordination inside the
complex unit cell of YAG \cite{Gracia}.

Fig.~\ref{Fig1} shows the auxiliary funtions $J(x/c;T)$ and $I(x/c)$
for octahedral symmetry, as given by Eqs.~(\ref{E3}) and (\ref{E4}).
Fig.~\ref{Fig2} represents the theoretically predicted interferogram

\begin{equation}
f(x;T)=\frac23 f_1(x;T)+\frac13 f_2(x;T)
\label{E11}
\end{equation}

\noindent
of two competing emissions with weights 2/3 and 1/3. The weighting
factors follow from assuming that the quasi--octahedrally coordinated
emission center has degenerate $x$ and $y$ lobes and slightly
different $z$ lobes. Both $f_1$ and $f_2$ have the functional form
(\ref{E10}) with constants $\alpha$, $\beta$ and $E$ chosen to fit the
measured spectrum. The transition energies are $E_1=2.535\,\rm{eV}$
and $E_2=2.343\,\rm{eV}$ (corresponding to the wavenumbers $\bar
k_1=E_1/(hc)=20450\,{\rm{cm}}^{-1}$ and $\bar k_2=E_2/(hc)
=18900\,{\rm{cm}}^{-1}$). The other constants are $\alpha_1=10.00$,
$\alpha_2=12.25$ and $\beta =\infty$ because $T\approx 0$. The value
of $E_1$ is given by the zero--phonon line, and hence is not an
adjustable parameter, the other three constants were chosen to fit the
experimental data.

Notice in Fig.~\ref{Fig2} that the number of the calculated points in
the interval of $x$ where $f(x/c;T)$ is appreciable seems insufficient
to represent properly the too structured function. This is not really
a problem to construct the spectrum because the FFT algorithm is an
analytic procedure that finds out the function whose Fourier transform
(or anti--transform) passes by the given points. The main aspect is
the accuracy of these points, but care must be taken also to avoid
aliasing of the frequency by a poor sampling of the data.

Fig.~(\ref{Fig3}) shows the cosine Fourier transform of the function
(\ref{E11}), calculated by the FFT algorithm included as a standard
tool in the Excel spreadsheet. The agreement of the theoretical curve
with the experimental spectrum is quite impressive. As the Fourier
transform is unique, this indicates that function (\ref{E11}) with the
assumed values for the constants should represent with good accuracy
the output of the Michelson interferometer. The FFT algorithm was run
with 1024 points for $x$, which runs over an interval of optical path
differences $0\le x\le 0.02\,\rm{mm}$.

\begin{figure}[h!]
\begin{center}
\includegraphics[width=10cm]{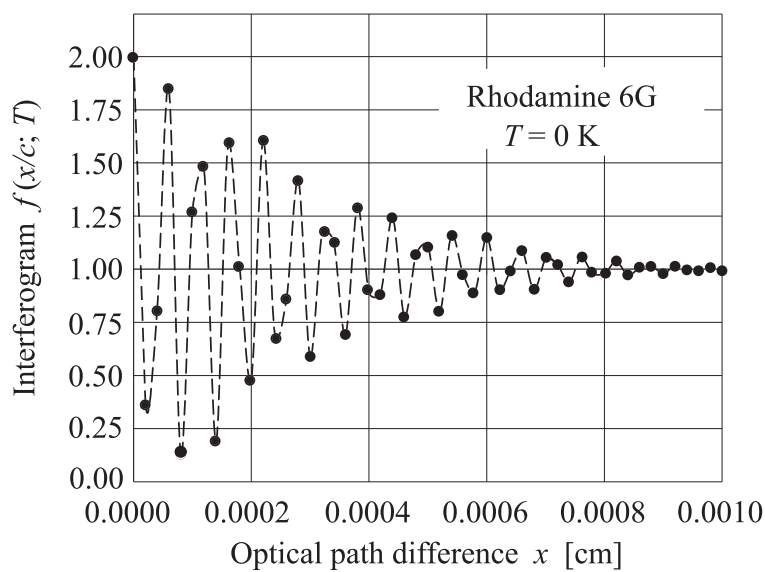}
\caption{\label{Fig4} Black dots represent function $f(x;T)$ given by
Eq.~(\ref{E12}), where $f_1$ and $f_2$ have the general form (\ref{E10})
with parameters $\alpha$ and $E$ chosen to fit the Michelson
interferogram of Rhodamine 6G. The broken line is only a guide to the
eye.}
\end{center}
\end{figure}

\begin{figure}[h!]
\begin{center}
\includegraphics[width=12cm]{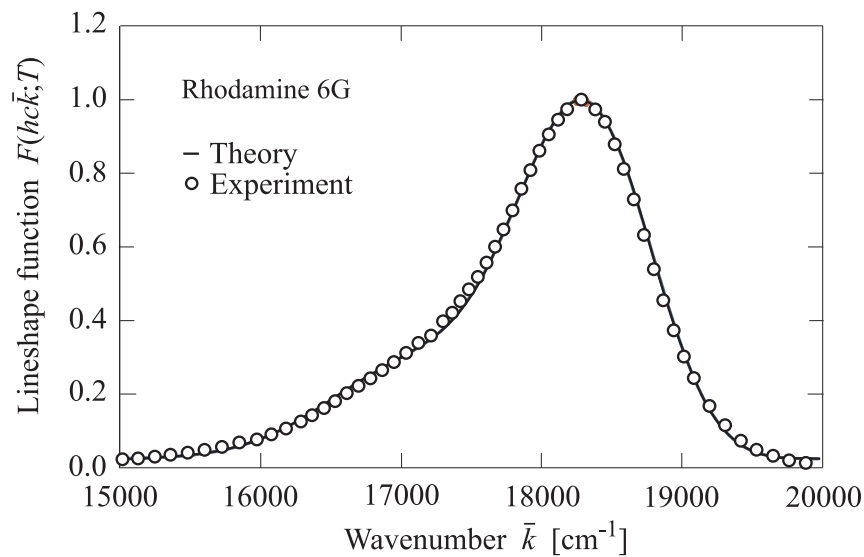}
\caption{\label{Fig5} Open circles represent the fluorescent emission
spectrum of Rhodamine 6G (Thermo Fisher Scientific, Fluorescence
SpectraViewer) and the solid line corresponds to the theoretical
lineshape function obtained from the Fourier transform of the function
passing by the black dots of Fig.~\ref{Fig4}.}
\end{center}
\end{figure}

Figure (\ref{Fig4}) shows a calculated interferogram of the general
form

\begin{equation}
f(x;T)=\frac34 f_1(x;T)+\frac14 f_2(x;T)
\label{E12}
\end{equation}

\noindent
where both $f_1$ and $f_2$ are given by Eq.~(\ref{E10}) with the
auxiliary functions (\ref{E7}) and (\ref{E8}) for tetrahedral
symmetry.  The weighting factors 3/4 and 1/4 assume a single distorted
orbital in the subgroup of coordination four inserted in the more
complex unit cell. The constants $\alpha_1 =18$ and $\alpha_2 =22$,
and the wavenumbers $\bar{k}_1 =20510\,\rm{cm}^{-1}$ and $\bar{k}_2
=19700\,\rm{cm}^{-1}$ ($E_1=2.543\,\rm{eV}$ and $E_1=2.442\,
\rm{eV}$), are chosen to give the fit of the experimental spectrum
of Rhodamine 6G shown in Fig.~\ref{Fig5}.

\section{Conclusions}

The method put forward here replaces the Fourier analysis of the
interferogram given as the output of a Fourier transform spectroscope
by a best fit analysis of Eq.~(\ref{E10}) to the data. The procedure
conserves all the advantages of traditional Fourier transform 
spectroscopy:\\
(i) Throughput (or Jacquinot) advantage. The energy throughput in the
interferometer is much higher than in a dispersive spectrometer because
has no slit selecting a narrow wavelength interval.\\ 
(ii) Multiplex (or Fellgett) advantage. The interferometer measures all
source wavelengths simultaneously and not by successive intervals, one
at a time, as in a dispersive instrument. (This and the former
attribute combine so that a Fourier transform spectrometer can achieve
much better signal-to-noise ratio than a dispersive instrument in a
shorter time).\\
(iii) Connes advantage. The scale for the optical path length
difference of the interferometer is given by a HeNe laser, which
provides a very fine, accurate and stable internal reference for each
scan. This is much more accurate and has much better long term
stability than the wavelength calibration of a dispersive instrument,
which is essentially mechanical.\\
(iv) Other advantages are the absence of random scattered light in the
Fourier transform spectrograph. The slit of a dispersive instrument
rejects most of the incoming light, which contributes to feed a diffuse
light level inside the apparatus. Also, the resolution of the
interferometer is constant at all wavenumbers, whereas the lower
throughput of dispersive instruments frequently obliges to adjust the
slit during the scan, modifying resolution.\\
But adds other important advantages:\\
(v) Numeric errors associated to the integration of rapidly oscillating
functions are avoided.\\ 
(vi) Apodization becomes unnecesary.\\
(vii) The procedure directly gives the net energy $E$ released or
captured by the electronic transition, with no Stokes shift.

\end{document}